\documentclass[aps,twocolumn,amsmath,amssymb,superscriptaddress,floatfix,nofootinbib]{revtex4}
\usepackage{graphicx}
\usepackage{epsfig}
\usepackage{epstopdf}
\usepackage{multirow}
\usepackage{color}
\usepackage[colorlinks=true,urlcolor=blue,linkcolor=blue,citecolor=blue]{hyperref}
\usepackage{hyperref}
\usepackage{amsmath}
\usepackage{amsfonts}
\usepackage{amssymb}
\usepackage{bbold}
\usepackage{braket}

\def \nn {\nonumber}
\def \) {\right)}

\begin{document}


\title{The Tjon Band in Nuclear Lattice Effective Field Theory}

\author{Nico~Klein} 
\affiliation{Helmholtz-Institut f\"ur Strahlen- und
             Kernphysik and Bethe Center for Theoretical Physics,
             Universit\"at Bonn,  D-53115 Bonn, Germany}

\author{Serdar~Elhatisari}
\affiliation{Helmholtz-Institut f\"ur Strahlen- und
             Kernphysik and Bethe Center for Theoretical Physics,
             Universit\"at Bonn,  D-53115 Bonn, Germany}  
\affiliation{Department of Physics, Karamanoglu Mehmetbey University, Karaman 70100, Turkey}

\author{Timo~A.~L\"{a}hde}
\affiliation{Institute~for~Advanced~Simulation, Institut~f\"{u}r~Kernphysik,
and J\"{u}lich~Center~for~Hadron~Physics,~Forschungszentrum~J\"{u}lich,
D-52425~J\"{u}lich, Germany}     

\author{Dean Lee}                 
\affiliation{National Superconducting Cyclotron Laboratory, Michigan State University, MI 48824, USA}
\affiliation{Department of Physics, North Carolina State University, Raleigh, NC 27695, USA}

\author{Ulf-G.~Mei{\ss}ner}  
\affiliation{Helmholtz-Institut f\"ur Strahlen- und
             Kernphysik and Bethe Center for Theoretical Physics,
             Universit\"at Bonn,  D-53115 Bonn, Germany}  
\affiliation{Institute~for~Advanced~Simulation, Institut~f\"{u}r~Kernphysik,
and J\"{u}lich~Center~for~Hadron~Physics,~Forschungszentrum~J\"{u}lich,
D-52425~J\"{u}lich, Germany}     
\affiliation{JARA~-~High~Performance~Computing, Forschungszentrum~J\"{u}lich,
D-52425 J\"{u}lich,~Germany} 
\date{\today}
\begin{abstract}
We explore the lattice spacing dependence in Nuclear Lattice Effective Field Theory 
for few-body systems up to next-to-next-to leading order in chiral effective field theory including all isospin breaking and 
electromagnetic effects, the complete two-pion-exchange potential and the three-nucleon forces. 
We calculate phase shifts in the neutron-proton system and proton-proton systems as well as the 
scattering length in the neutron-neutron system. We then perform a full next-to-next-to-leading order
calculation with two-nucleon and three-nucleon forces for the triton and helium-4 and analyse their 
binding energy correlation. We show how the Tjon band is reached by decreasing the lattice spacing 
and confirm the continuum observation that a four-body force is not necessary to describe light nuclei. 
\end{abstract}

\maketitle
\section{Introduction}
Nuclear Lattice Effective Field Theory (NLEFT) has become a powerful tool in the last years to study 
the formation of nuclei from nucleons in an \textit{ab initio} way. Using this method it was possible 
to calculate the binding energies of medium mass nuclei with good accuracy \cite{Epelbaum:2009pd,Lahde:2013uqa} and 
postdict the Hoyle state \cite{Epelbaum:2011md,Epelbaum:2012qn}, which is an excited state in carbon-12 indispensable 
for nucleosynthesis in stars. Besides binding energies, also scattering processes like nucleon-nucleon 
\cite{Alarcon:2017zcv, Borasoy:2007vy} or alpha-alpha-scattering \cite{Elhatisari:2015iga} were investigated. 
NLEFT combines two powerful concepts.
First, we have chiral nuclear effective field theory \cite{Epelbaum:2005pn, Epelbaum:2008ga}, which 
gives a systematic description of low-energy hadron physics based on the symmetries (and their
breaking) of the underlying gauge field 
theory, Quantum Chromodynamics. This continuum approach can be combined with well established many-body 
continuum schemes to go beyond light nuclei, such as the shell model, the no-core-shell model, coupled 
cluster theory, variational Monte Carlo methods, and so on, 
see e.g.~Refs.~\cite{Navratil:2007we,Hagen:2008iw,Otsuka:2009cs,Holt:2011fj,Binder:2012mk,Soma:2013xha,Wienholtz:2013nya,Binder:2015mbz,Ekstrom:2015rta,Lynn:2017fxg,Binder:2018pgl}.
Second, we discretize Euclidean space time with spatial lattice spacing $a$ and temporal lattice spacing $a_t$ and 
use Monte Carlo methods for the numerical evaluation of few- and many-body problems. While many interesting and precise results could be obtained in this scheme, 
a problem due to the discretization of space had  never been resolved in a satisfactory fashion. 
In Ref.~\cite{Epelbaum:2009pd} an overbinding of the ground state of $^4$He was observed which was traced 
back to the appearance of implicit multi-particle interactions which have significant effect on 
few-body physics on coarse lattices. In Ref.~\cite{Lee:2005nm}, this effect was demonstrated explicitly for an
N-boson system in two dimensions with short-range interaction. 
This effect becomes even stronger in larger nuclei. In Ref.~\cite{Lahde:2013uqa} this discretization artefact was 
cured by adding an {\em effective} four-nucleon force (4NF), that, however, is not related to the 
chiral expansion. Here, we want to reconsider this issue and prove the conjecture of the appearance of the implicit multi-particle interactions and show with decreasing lattice spacing their effect is diminished more and more. Consequently, the effective 4NF is not necessary any more once the lattice spacing is chosen small enough, and one should also confirm the correlation between 3N and 4N systems, known as the Tjon 
band \cite{Tjon:1975sme,Platter:2004zs}. Note that historically, this correlation was called the Tjon line, but as
theory has an inherent uncertainty, it really is a band as stressed in Ref.~\cite{Platter:2004zs}. 
While the aforementioned overbinding of $^4$He on coarse 
lattices results in correlation points relatively far off the Tjon band, the binding energies for 
triton and helium-4 should get closer to this band or already be on top of it once the lattice spacing is 
chosen small enough. Furthermore, note that one also has to be aware of an additional effect that
needs proper treatment. The configurations with four nucleons on one lattice site require smearing
as otherwise a strong overbinding due to these is generated~\cite{Borasoy:2006qn}. 
In the chiral EFT action used for most investigations, this smearing was adjusted to get the proper 
neutron-proton effective range. This procedure
might, however, not be sufficient in larger systems, as evidenced by the highly successful non-locally smeared
leading order action proposed in Ref.~\cite{Elhatisari:2017eno}.

The EFT provides a counting scheme for the expansion of the effective potential \cite{Weinberg:1991um}
systematically up to any given order $\mathcal{O}(Q/\Lambda)$, where $Q$ is a small expansion 
parameter with respect to the nuclear hard scale $\Lambda\approx 500\,$MeV. Note that this hard 
scale is smaller than the usual chiral perturbation theory scale because of the non-perturbative nature of the nuclear interactions. For a detailed discussion, see e.g.~Ref.~\cite{Epelbaum:2014efa}.
The pertinent small expansion parameter in our case is the 
nucleon momentum $p$ or the pion mass $M_\pi$ or the electromagnetic charge $e\propto M_\pi/\Lambda$. 
Within this counting scheme, the two-body contact interactions start at leading order (LO), 
$\mathcal{O}[(Q/\Lambda)^0]$ while momentum-dependent and electromagnetic interactions are at 
next-to-leading order (NLO), $\mathcal{O}[(Q/\Lambda)^2]$ and do not have any additional contribution at 
N2LO, $\mathcal{O}[(Q/\Lambda)^3]$. An additional contact contribution would arise at N3LO, 
$\mathcal{O}[(Q/\Lambda)^4]$, which is beyond the accuracy of our calculations in the three- and 
four-body sector performed here. The two-pion-exchange potential (TPEP) has contributions at NLO and N2LO. 
A detailed analysis of the lattice space dependence of the two-body sector can be found in 
Ref.~\cite{Alarcon:2017zcv}, where higher-order corrections were included both perturbatively and 
non-perturbatively. In the following we will include and extend the two-body analysis, but we will focus 
on the perturbative approach as we want to be in agreement three- and four-body calculations in which
all corrections beyond LO have been included perturbatively. In the three-body sector, 3NF corrections 
only start at N2LO as NLO contributions only consist of reducible topologies which do not produce any 
non-vanishing contributions. Concerning the 4NF, it was conjectured that they are not necessary 
due to the same argument of vanishing contributions for 3NF. However, it was shown by Ref.~\cite{Epelbaum:2007us} 
that these forces matter at N3LO and a rough estimation of some of their contribution gives to the binding energy of $^4$He is about $100$~keV \cite{nogga}. Hence, they are 
are beyond the order we include in our analysis and 
their actual contribution is beyond the accuracy of our work as well, so we can safely neglect them.
We remark that these chiral 4NFs are not the effective 4NFs that were included in Ref.~\cite{Lahde:2013uqa}.

The paper is organized as follows. In Sec.~\ref{sec:twobody} we introduce the method of 
NLEFT with an emphasis on the two-body sector. We calculate phase shifts up to N2LO for neutron-proton 
scattering and we also consider electromagnetic corrections for neutron-neutron and proton-proton scattering. 
Then in Sec.~\ref{sec:threebody}, we extend the method to three-body systems and calculate the properties of 
triton at each order in the framework of NLEFT. When describing the four-body system in Sec.~\ref{sec:fourbody}, 
we also give a brief introduction into Monte Carlo simulations which are necessary for the calculation of the
$^4$He properties. All calculations are done for lattice spacings of $a=1.97$~fm, $a=1.64$~fm and $a=1.32$~fm, 
which means that the respective cutoff $\Lambda_a=\pi/a$ remains below the breakdown scale of the theory. 
The respective temporal lattice spacing is chosen as $a_t=1.32$~fm, $a_t=0.91$~fm and $a_t=0.59$~fm, such that the ratio $a^2/a_t$ is kept fixed. 
Finally, we investigate the $^3$H-$^4$He correlation in Sec.~\ref{sec:Tjon}. In Sec.~\ref{sec:conclusions} 
we conclude and give an outlook on further improvements.

\vfill

\section{Two-body-sector}\label{sec:twobody}
\subsection{Theoretical framework}

In the two-body sector we solve the LO non-pertubatively and we include the NLO and N2LO 
corrections perturbatively.
For the free part of the Hamiltonian, we use an $\mathcal{O}(a^4)$-improved version
\begin{equation}
\begin{split}
&H_{\rm free}=\frac{1}{2 m_N}\sum_{\vec{n},i,j}\sum_{\hat{s}}2 \omega_0 a^\dagger_{i,j}(\vec{n})a_{i,j}(\vec{n}) \\ 
&+\sum_{l=1}^3(-1)^k\omega_l[a_{i,j}^\dagger(\vec{n}a_{i,j})(\vec{n}+l\hat{s})+a_{i,j}^\dagger(\vec{n})a_{i,j}(\vec{n}-l\hat{s})]~.
\label{eq:Hfree}
\end{split}
\end{equation}
The coefficients $\omega_l$ represent the improved action including a stretching factor which connects $\mathcal{O}(a^4)$ action with $\mathcal{O}(a^3)$ action to correct the dispersion relation of the nucleon-nucleon system. Explicitly we use 
\begin{align}
\omega_0&= 10\left(\frac{49}{36}-\frac{5}{4}\right)+\frac{49}{36},  &\omega_1 = 
10\left(\frac{3}{2}-\frac{4}{3}\right)+\frac{3}{2},\;\;\;\nonumber\\
\omega_2 &= 10\left(\frac{3}{20}-\frac{1}{12}\right)+\frac{3}{20},  &\omega_3 =10\left(\frac{1}{90}-0\right)+\frac{1}{90}.\nonumber\\
\end{align}
For details on this, see Ref.~\cite{Lee:2008fa}.
In the LO potential, we include smeared contact interaction operators which are projected on the S-waves and 
the one-pion-exchange (OPE) potential. The short-range contact interaction reads
\begin{equation}
\begin{split}
H&_{\rm LO,contact}=\sum_{\vec{n}_1,\vec{n}_2}f\left(\vec{n}_1-\vec{n}_2\right)\colon \left[c_0\rho^{a^\dagger,a}(\vec{n}_1)\rho^{a^\dagger,a}(\vec{n}_2)\right.\\
+&c_{ss}\sum_{S=1}^3\rho^{a^\dagger,a}_S(\vec{n}_1)\rho^{a^\dagger,a}_S(\vec{n}_2)
+c_{ii}\sum_{I=1}^3\rho^{a^\dagger,a}_I(\vec{n}_1)\rho^{a^\dagger,a}_I(\vec{n}_2)\\
+&\left.c_{si}\sum_{S,I=1}^3\rho^{a^\dagger,a}_{S,I}(\vec{n}_1)\rho^{a^\dagger,a}_{S,I}(\vec{n}_2)\right] \colon \\
\end{split}
\label{eq:HLOcontact}
\end{equation}
with 
\begin{eqnarray}
c_0 &=& (3C_{^1S_0}+3C_{^3S_1})/16~,~c_{ss}=(-3C_{^1S_0}+C_{^3S_1})/16~,\nonumber\\
c_{ii}&=& (C_{^1S_0}-3C_{^3S_1})/16~,~~~c_{is}=(-C_{^1S_0}-C_{^3S_1})/16~.\nonumber\\
\end{eqnarray}
$f(\vec{n})$ is the so-called smearing function, defined 
by 
\begin{equation}
f(\vec{n})=\mathcal{F}[f_0^{-1}\exp(-b \vec{q}\,^4/4)](\vec{n})~, 
\end{equation}
with $\mathcal{F}$ denoting the
Fourier transformation in the discrete space, and $b$ is the smearing parameter.
 Furthermore, the normalization constant is given by
$f_0=\sum_{\vec{q}}\exp(-b\vec{q}\,^4/4)$. For the momentum-squared discretization we use also 
an $\mathcal{O}(a^4)$ improved one, 
\begin{equation}
\vec{q}\,^2  = 6\omega_0+2\sum_{l=1}^3\sum_{s=1}^3(-1)^n\omega_s\cos\left(\frac{2 s \pi k_l}{L}\right)~.
 \label{eq:q2}
\end{equation} 
As already argued in Ref.~\cite{Borasoy:2006qn} this smearing improves the S-wave description above a relative momentum of 50 MeV and hence reduces the clustering instability for the few-body systems like $^4$He. The long-range OPE is given by 
\begin{equation}
\begin{split}
H_{\rm OPE}&=-\frac{g_A^2}{8F_\pi^2}\sum_{S_1,S_2,I}\sum_{n_1,n_2}G_{S_1,S_2}(\vec{n}_1-\vec{n}_2)\\
&\times \colon \rho^{a^\dagger,a}_{S_1,I}(\vec{n}_1)\rho^{a^\dagger,a}_{S_2,I}(\vec{n}_2) \colon
\end{split}
\label{eq:HOPE}
\end{equation}
where the pion propagator is given by
\begin{equation}
G_{S_1,S_2}\left(\vec{q}\,\right)=\frac{q_{S_1}q_{S_2}}{M_\pi^2+\vec{q}\,^{2}}
\end{equation}
with
\begin{equation}
q_S=\sin\left(\frac{2\pi k_S}{L}\right), \label{eq:qs}
\end{equation}
in momentum space. For the pion-nucleon coupling constant and the 
pion decay constant, we use the values $g_A=1.29$ (to account for the Goldberger-Treiman discrepancy)
and $F_\pi = 92.2$~MeV.
While there are various ways to regularize the long-range part with a Gaussian smearing function 
in momentum space or an analogue function directly in position space \cite{Epelbaum:2014efa}, 
a similar analysis was done in \cite{Klein:2015vna} for the NLEFT approach where it was argued that 
the hard cut-off regularization due to lattice spacing is sufficient up to those we are 
interested in.

The NLO contact interaction contribution consists of ten terms totally. First, we have to 
include two operators which are not NLO operators by power counting but which are necessary 
for LO corrections. More precisely, for a coarse lattice spacing, these represent the full
non-local structure of the TPEP \cite{Epelbaum:2010xt} while for finer lattices they could
be dropped. However, we keep them for consistency. These terms read:
\begin{eqnarray}
V_{\rm NLO}^1 &=& -\frac{\Delta C}{2} :\sum_{\vec n} \rho(\vec{n}) \rho(\vec{n}):,  \label{NLO_1}\\
V_{\rm NLO}^2 &=& -\frac{\Delta C_{I^2}}{2} :\sum_{\vec n} \sum_{I=1}^3 \rho_I(\vec{n})  \rho_I(\vec{n}):.  \label{NLO_2}
\end{eqnarray}
Then we have the seven standard NLO contact interactions, 
\begin{eqnarray}
V_{\rm NLO}^3 &=& -\frac{C_{q^2}}{2} :\sum_{\vec n} \sum_{l=1}^3 \rho(\vec{n}) \nabla_{l}^2 \rho(\vec{n}):,  \label{NLO_i}\\
V_{\rm NLO}^4 &=& -\frac{C_{I^2,q^2}}{2} :\sum_{\vec{n}} \sum_{I=1}^3 \sum_{l=1}^3 \rho_I^{}(\vec{n}) \nabla_l^2 \rho_I^{}(\vec{n}):,  \\
V_{\rm NLO}^5 &=& -\frac{C_{S^2,q^2}}{2} :\sum_{\vec{n}} \sum_{S=1}^3 \sum_{l=1}^3 \rho_S^{}(\vec{n}) \nabla_l^2\rho_S^{}(\vec{n}):, \\
V_{\rm NLO}^6 &=& -\frac{C_{S^2,I^2,q^2}}{2}   \nonumber\\
&& \times  :\sum_{\vec{n}}\sum_{S,I=1}^3 \rho_{S, I}^{}(\vec{n}) 
\nabla_l^2 \rho_{S, I}^{}(\vec{n}):,  \label{projection_NLO} \\  
V_{\rm NLO}^7 &=& \frac{C_{(q\cdot S)^2}}{2} :\sum_{\vec{n}} \sum_{S=1}^3 \nabla_S^{} \rho_{S}^{}(\vec{n})\nonumber\\ 
&& \times  \sum_{S^\prime=1}^3 \nabla_{S^\prime}^{} \rho_{S^\prime}^{}(\vec{n}):, \\
V_{\rm NLO}^8 &=& \frac{C_{I^2,(q\cdot S)^2}}{2} :\sum_{\vec{n}} \sum_{S=1}^3 \nabla_S^{} \rho_{S, I}^{}(\vec{n}) \\
&&\times \sum_{S^\prime=1}^3 \nabla_{S^\prime}^{} \rho_{S^\prime}^{} (\vec{n}):, \label{NLO_j} \nonumber\\
V_{\rm NLO}^9 &=& -\frac{iC^{I=1}_{(q\times S)\cdot k}}{2} : \sum_{\vec{n}} \sum_{S=1}^3 \sum_{l, l^\prime = 1}^3 \varepsilon_{l, S, l^\prime} \\
&& \times 
\Bigg[\Pi_l^{}(\vec{n}) \nabla_{l^\prime}^{} \rho_S^{} (\vec{n}) \nn + \: \Pi_{l, S}^{}(\vec{n}) \nabla_{l^\prime}^{}
\rho(\vec{n}) \Bigg]:~,
\end{eqnarray}
with $\varepsilon_{a,b,c}$ the totally antisymmetric Levi-Civita tensor in three dimensions.
Finally, we also include the following SO(3) breaking term
\begin{equation}
V_{\rm NLO}^{10} = \frac{C_{SO(3)}}{2}:\sum_S \rho_{S}^{}(\vec{n})\nabla_S^2\rho_{S}^{}(\vec{n}): \label{eq:SO3breaking},
\end{equation}
which allows us to remove lattice artefacts (unphysical partial wave mixing) due to rotational 
symmetry breaking. More specifically, this term is tuned to remove the mixing between the $^3S_1$-$^3D_1$ and
the  $^3D_3$-$^3G_3$ channels. Unphysical partial wave mixing in higher waves is so small that it can be
ignored. More details of the notation can be found App.~\ref{app_operators}. 

At N2LO, 
there are no further contact terms. Further, we need to include the TPEP at NLO and N2LO. 
While they are largely absorbed in the NLO contact terms for a very coarse lattice, 
they play an important role for finer lattices \cite{Epelbaum:2009pd}.
At NLO, the TPEP reads


\begin{equation}
\begin{split}
&V_{\rm NLO}^{\rm TPEP} =
\sum_{\vec{n}_1,\vec{n}_2}\left\{\sum_{I=1}^3-\frac{\colon
\rho_I\left(\vec{n}_1\right)\rho_I\left(\vec{n}_2\right)\colon
}{384\pi^2F_\pi^4}\right.\\
&\times\left[4M_\pi^2\left(5g_A^4-4g_A^2-1\right)V_{\rm NLO}^{\rm
TPEP,1}(\vec{n}_1-\vec{n}_2)\right. \\
& \left. +\left(23g_A^4-10g_A^2-1\right)V_{\rm NLO}^{\rm
TPEP,2}(\vec{n}_1-\vec{n}_2)\right.\\
&\left.+48g_A^4M_\pi^4V_{\rm NLO}^{\rm
TPEP,3}(\vec{n}_1-\vec{n}_2)\right]-\frac{3g_A^4}{64\pi^2F_\pi^4} \\
&\times \left[\colon \sum_{S_1,S_2=1}^3V_{\rm NLO}^{\rm
TPEP,4}(\vec{n}_1-\vec{n}_2,S_1,S_2)
\rho_{S_1}\left(\vec{n}_1\right)\rho_{S_2}\left(\vec{n}_2\right) \right.\\
& \left. \left.-\sum_{S=1}^3 V_{\rm NLO}^{\rm
TPEP,2}(\vec{n}_1-\vec{n}_2)
\rho_S\left(\vec{n}_1\right)\rho_S\left(\vec{n}_2\right) \colon
\right]\right\}
\end{split}
\end{equation}
with the Fourier-transformed parts
\begin{align}
&V_{\rm NLO}^{\rm TPEP,1}(\vec{n})=\mathcal{F}\left[L\left(\lvert
\vec{q}\, \rvert\right)\right]\left(\vec{n}\right),\\
&V_{\rm NLO}^{\rm TPEP,2}(\vec{n})=\mathcal{F}\left[L\left(\lvert
\vec{q}\, \rvert\right)\vec{q}\,^2\,\right]\left(\vec{n}\right),\\
&V_{\rm NLO}^{\rm TPEP,3}(\vec{n})=\mathcal{F}\left[L\left(\lvert
\vec{q}\,
\rvert\right)\vec{q}\,^2\frac{1}{4M_\pi^2+\vec{q}\,^2}\right]\left(\vec{n}\right),\\
&V_{\rm NLO}^{\rm
TPEP,4}(\vec{n},S_1,S_2)=\mathcal{F}\left[L\left(\lvert \vec{q}\,
\rvert\right)q_{S_1}q_{S_2}\right]\left(\vec{n}\right),
\end{align}
where
\begin{equation}
L\left(q\right)=\frac{1}{2q}\sqrt{4M_\pi^2+q^2}\log\frac{\sqrt{4M_\pi^2+q^2}+q}{\sqrt{4M_\pi^2+q^2}-q},
\end{equation}
and $\lvert \vec{q}\, \rvert=\sqrt{\sum_{i=1}^3 q_i^2}$. Note that this notation is different from Eq.~(\ref{eq:q2}), 
as the TPEP contributions should be absorbed in the NLO contact interaction up to $\mathcal{O}(Q^2)$.

At N2LO, the TPE potential has a subleading contribution given by
\begin{equation}
\begin{split}
&V_{\rm N2LO}^{\rm TPEP} = \sum_{\vec{n}_1,\vec{n}_2}\left\{-\frac{3g_A^2}{16\pi F_\pi^4}\colon \rho\left(\vec{n}_1\right)\rho\left(\vec{n}_2\right)\colon \right.\\
&\times\left[2M_\pi^2\left(2c_1-c_3\right)V_{\rm N2LO}^{\rm TPEP,1}(\vec{n_1}-\vec{n}_2)\right.\\
&\qquad\left.-c_3 V_{\rm N2LO}^{\rm TPEP,2}(\vec{n_1}-\vec{n}_2)\right]-\frac{g_A^4c_4}{32\pi F_\pi^4}\\
&\times \left[\colon \sum_{S_1,S_2=1}^3\rho_{S_1}\left(\vec{n}_1\right)\rho_{S_2}\left(\vec{n}_2\right)
V_{\rm N2LO}^{\rm TPEP,3}(\vec{n_1}-\vec{n}_2,S_1,S_2)\right.\\
&\left. \left.-\sum_{S=1}^3\rho_S\left(\vec{n}_1\right)\rho_S\left(\vec{n}_2\right)
V_{\rm N2LO}^{\rm TPEP,4}(\vec{n_1}-\vec{n}_2)\colon \right]\right\}
\end{split}
\end{equation}
with $c_1=-1.10$~GeV$^{-1}$, $c_3=-5.54$~GeV$^{-1}$ and $c_4=4.17$~GeV$^{-1}$ \cite{Hoferichter:2015tha} and
\begin{equation}
A\left(q\right)=\frac{1}{2q} \arctan\frac{q}{2M_\pi}
\end{equation}
with
\begin{align}
&V_{\rm N2LO}^{\rm TPEP,1}(\vec{n})=\mathcal{F}\left[A(\lvert \vec{q}\, \rvert)(2M_\pi^2+\vec{q}\,^2) \right](\vec{n}),\\
&V_{\rm N2LO}^{\rm TPEP,2}(\vec{n})=\mathcal{F}\left[A(\lvert \vec{q}\, \rvert)\left(2M_\pi^2+\vec{q}\,^2\right)\vec{q}\,^2  \right](\vec{n}),\\
\begin{split}
&V_{\rm N2LO}^{\rm TPEP,3}(\vec{n},S_1,S_2)=\\
& \qquad \mathcal{F}\left[A(\lvert \vec{q}\, \rvert)(4M_\pi^2+\vec{q}\,^2)q_{S_1}q_{S_2} \right](\vec{n},S_1,S_2),
\end{split}\\
&V_{\rm N2LO}^{\rm TPEP,4}(\vec{n})=\mathcal{F}\left[A(\lvert \vec{q}\, \rvert)(4M_\pi^2+\vec{q}\,^2)\vec{q}\,^2 \right](\vec{n}).
\end{align}
Additionally, we also include isospin-breaking effects due to the different pion masses and 
corrections due to an improved version of the OPE. The correction is given by 
\begin{equation}
\begin{split}
V_{\rm Dx}&=-\frac{g_A^2}{8F_\pi^2}\sum_{S_1,S_2,I}\sum_{\vec{n}_1,\vec{n}_2}\left[\tilde{G}_{S_1,S_2}(\vec{n}_1-\vec{n}_2)  \right.\\
&-\left.G_{S_1,S_2}(\vec{n}_1-\vec{n}_2)\right]\colon\rho^{a^\dagger,a}_{S_1,I}(\vec{n}_1)\rho^{a^\dagger,a}_{S_2,I}(\vec{n}_2)\colon,
\end{split}
\end{equation}
where the improved propagator is defined as
\begin{equation}
\tilde{G}_{S_1,S_2}\left(\vec{q}\,\right)=\frac{\tilde{q}_{S_1}\tilde{q}_{S_2}}{M_\pi^2+\vec{q}\,^{2}},
\end{equation}
with $\tilde{q}_S=(4/3) \sin(2\pi k/L)+ (1/6)\sin(4\pi k/L)$ and $\vec{q}\,^2$ according to Eq.~(\ref{eq:q2}).\\
The isospin corrections caused by the pion mass differences are defined as
\begin{equation}
\begin{split}
V_{\rm IB}&=-\frac{g_A^2}{8F_\pi^2}\sum_{S_1,S_2,I}\sum_{\vec{n}_1,\vec{n}_2}\left[\bar{G}_{S_1,S_2}(\vec{n}_1-\vec{n}_2)  \right.\\
&-\left.G_{S_1,S_2}(\vec{n}_1-\vec{n}_2)\right]\colon\rho^{a^\dagger,a}_{S_1,I}(\vec{n}_1)\rho^{a^\dagger,a}_{S_2,I}(\vec{n}_2)\colon~.
\end{split}
\end{equation}
 The pion propagator with charged pions reads
\begin{equation}
\bar{G}_{S_1,S_2}\left(\vec{q}\,\right)=\frac{q_{S_1}q_{S_2}}{M_{\pi_\pm}^{2}+\vec{q}\,^{2}}.
\end{equation}
Note that since isospin breaking is an NLO correction, we do not need to include the
corrections for the charged pion propagator as just discussed for the neutral pion one.
Hence, the complete 2N N2LO Hamiltonian reads
\begin{equation}
H_{\rm N2LO}^{\rm 2N} = \sum_{i=1}^{10} V_{\rm NLO}^i+V_{\rm NLO}^{\rm TPEP}+V_{\rm N2LO}^{\rm TPEP}+V_{\rm Dx}+V_{\rm IB}
\end{equation}

As we want to describe light nuclei in a later stage, we also have to include Coulomb forces 
as well as proton-proton and neutron-neutron contact terms (for details, see Ref.~\cite{Epelbaum:2010xt})
\begin{align}
V^{\rm Coul}&=\frac{\alpha_{\rm EM}}{2}\sum_{\vec{n}_1, \vec{n}_2}\frac{1}{\mathrm{max}\left(0.5,\lvert\vec{n}_1-\vec{n}_2 \rvert\right)} \colon \rho_p(\vec{n}_1)\rho_p(\vec{n}_2)\colon ,\\
V^{pp}&= \frac{C_{pp}}{2} \sum_{n}\colon \rho_p(\vec{n})\rho_p(\vec{n})\colon ,\\
V^{nn}&=\frac{C_{nn}}{2} \sum_{n}\colon \rho_n(\vec{n})\rho_n(\vec{n})\colon .
\end{align}
with $\alpha_{\rm EM}$ the electromagnetic fine-structure constant and the projection densities 
are given in App.~\ref{app_operators}. Thus, the complete electromagnetic contribution reads
\begin{equation}
V_{\rm EM}^{\rm 2N}=V^{\rm Coul}+V^{pp}+V^{nn}.
\end{equation}
The alert reader might notice  that $V^{pp}$ and $V^{nn}$ are really strong isospin-breaking terms. We book them
here, because $V^{pp}$ is used to renormalize the Coulomb potential.

Now, the Hamiltonian is defined and, in the standard approach, we introduce a spherical wall boundary on the relative separation between nucleons in order to compute scattering phase shifts. This spherical wall is placed at radius $R_{\rm wall}$ outside the interaction region. We then solve for standing waves solutions of the transfer matrix \cite{Lee:2008fa}
\begin{equation}
\colon\exp\left[-\alpha_t\left(H_{\rm free}+H_{\rm LO}\right)\right]\colon \ket{\Psi}= \lambda \ket{\Psi},
\label{eq:eigvaleq}
\end{equation}
with $\alpha_t=a_t/a$ and the energy given by
\begin{equation}
E=-\frac{\log\left(\lambda\right)}{\alpha_t}~.
\label{eq:EBextract}
\end{equation}
 The solutions must be identified with the correct partial wave, then one could use the 
energy shift between the free system without any interaction and the one with interaction 
to calculate the phase shifts \cite{Borasoy:2007vy}. 
The NLO and N2LO energy corrections are implemented perturbatively just by calculating 
the corresponding matrix element. In what follows, we utilize a more sosphisticated procedure: 
Using the radial projection method we impose a spherical wall, but we first project the system onto 
its partial waves where the only degree of freedom is the radial one. This projection 
accelerates the fit procedures and is necessary particularly in the case of small lattice spacings.
Then the basis turns from a three-dimensional vector $\ket{\vec{R}}$ to a radial basis $\ket{R}$:
\begin{equation}
\ket{R} = \sum_{\vec{R}^\prime} Y_{l,l_z}(\hat{R}^\prime)\delta_{R,R^\prime}\ket{\vec{R}^\prime}.
\end{equation}
Here, the $Y_{l,l_z}$ are the spherical harmonics specified by their angular momentum $l,l_z$. 
Consequently, all operators are projected to a radial basis, too:
\begin{equation}
\mathcal{O}(\vec{R}) \rightarrow \mathcal{O}\left(R\right) 
\end{equation}
and the problem can be solved completely in the reduced basis analoguous to 
Eqs.~(\ref{eq:eigvaleq},\ref{eq:EBextract}). Details of the method can be found in 
Refs.~\cite{Lu:2015riz,Elhatisari:2016hby} 
where the projection, the binning, the new radial metric and the extraction method for coupled 
channels are explained.
In an uncoupled channel, the projected radial wave function solution $\psi_l^p(r)$ can be directly 
identified with the spherical Bessel functions in a region between the interaction region and 
the spherical wall, which we confine to be between $R_{\rm in}$ and $R_{\rm out}$. Thus, the phase 
shift $\delta_l$ can be read off immediately,
\begin{equation}
\psi_l^p(r)=\mathcal{N}^p\left[\cot \left(\delta_l\right) j_l(pr)+ n_l(pr)\right],
\end{equation}
where $p$ is the relative momentum, $\mathcal{N}^p$ a normalization constant and $j_l(pr)$ and $n_l(pr)$ the spherical Bessel functions of first and second kind.
For the perturbative energy corrections, we have to use the projected potentials and the new 
phase shifts are calculated using the energy shifts according to Ref.~\cite{Borasoy:2007vy}. For 
the lattice spacing of $a=1.97$~fm we use $L = 32$~[l.u.], $R_{\rm wall}=14.02$~[l.u.], 
$R_{\rm in}=9.02$~and $R_{\rm out}=12.02$~[l.u.]. For the smaller lattice spacings, we use the 
same values for these parameters in physical units (fm).
For the neutron-proton fit procedure we follow Ref.~\cite{Alarcon:2017zcv}. In general, we do a 
$\chi^2$ fit to partial wave analysis data PWA, NijmI, NijmII and Reid93 of \cite{Stoks:1994wp} 
according to \cite{Epelbaum:2014efa},
\begin{equation}
\chi^2=\sum_i\frac{\left(\delta_i-\delta_i^{\mathrm{PWA}}\right)^2}{\Delta_i^2}\label{chisquare}
\end{equation}
where the error is defined as $\Delta_i=\mathrm{max}[\Delta_i^{\mathrm{PWA}},\lvert \delta_i^{\mathrm{NijmI}}
-\delta_i^{\mathrm{PWA}}\rvert,\lvert \delta_i^{\mathrm{NijmII}}-\delta_i^{\mathrm{PWA}}\rvert,\lvert 
\delta_i^{\mathrm{Reid93}}-\delta_i^{\mathrm{PWA}}\rvert ]$. Further details on errors and error 
propagation can be found in App.~\ref{app_errors}.
At LO we fit the LECs $C_{^1S_0}$ and $C_{^3S_1}$ to the $^1S_0$ and $^3S_1$ channel up to $100$~MeV, and
we keep the smearing parameter $b$ fixed at $b=0.07$. While we could use it as a fit parameter 
as well for the coarse lattice, it will cause some problem for small lattice spacing. As 
one can see in Ref.~\cite{Alarcon:2017zcv}, most of the partial waves are better described
with smaller lattice spacing except for the $^1S_0$ channel which becomes too strong. 
The reason is that the LO smearing parameter is mainly determined by the $^3S_1$ channel due to 
the different errors in the PWA analysis. This effect is negligble for large lattice spacings but 
becomes sizeable for smaller ones and worsens the prediction of the $^1S_0$ wave. Hence, we keep 
the smearing parameter close to the fit value for $a=1.97$~fm and all corrections are done 
by NLO and N2LO insertions.

Once the LO is fixed, we include isospin-breaking effects, the improved description of the 
OPEP, the TPEP at NLO and N2LO as well as the NLO contact terms. We fit all remaining coefficients 
to S- and P-waves up to $150$~MeV momentum as well as the deuteron binding energy.
Afterwards we fit the proton-proton interaction term to the pp $^1S_0$ phase shift and the 
neutron-neutron interaction term to the nn-scattering length of $a_{nn}=18(1)$~fm.
Due to the long-range nature of the Coulomb force we include it non-perturbatively in the pp channel and change the Bessel functions with the respective Coulomb ones \cite{Epelbaum:2009pd}, namely $j_l(pr)$ by $F_l(\eta,pr)$ and $n_l(pr)$ by $G_l(\eta,pr)$, where $\eta=\alpha_{\rm EM}m/(2p)$ and
\begin{align}
\begin{split}
F_l(\eta,pr)&=(pr)^{l+1}\exp(-i pr)c_l(\eta)\\
\times & _1F_1(l+1-i\eta,2l+2,2ipr),
\end{split}\\
\begin{split}
G_l(\eta,pr)&=\frac{(2i)^{2l+1}(pr)^{l+1}\exp(-ipr)\Gamma(l+1-i\eta)}{\Gamma(2l+2)c_l(\eta)}\\
 \times  & U(l+1-i\eta,2l+2,2ipr)+iF_l(\eta,pr),
\end{split}
\end{align}
with $_1F_1$ and $U$ the Kummer functions of the first and second kind while $c_l$ is defined as
\begin{equation}
c_l(\eta)=\frac{2^l\exp(-\pi\eta/2)\lvert\Gamma(l+1+i\eta)}{\Gamma(2l+2)}.
\end{equation}
However, the contact interaction $V_{pp}$ is included perturbatively as all other higher-order operators.
\subsection{Results}
\begin{figure*}[hbt]
\begin{center}
\includegraphics[width=\textwidth]{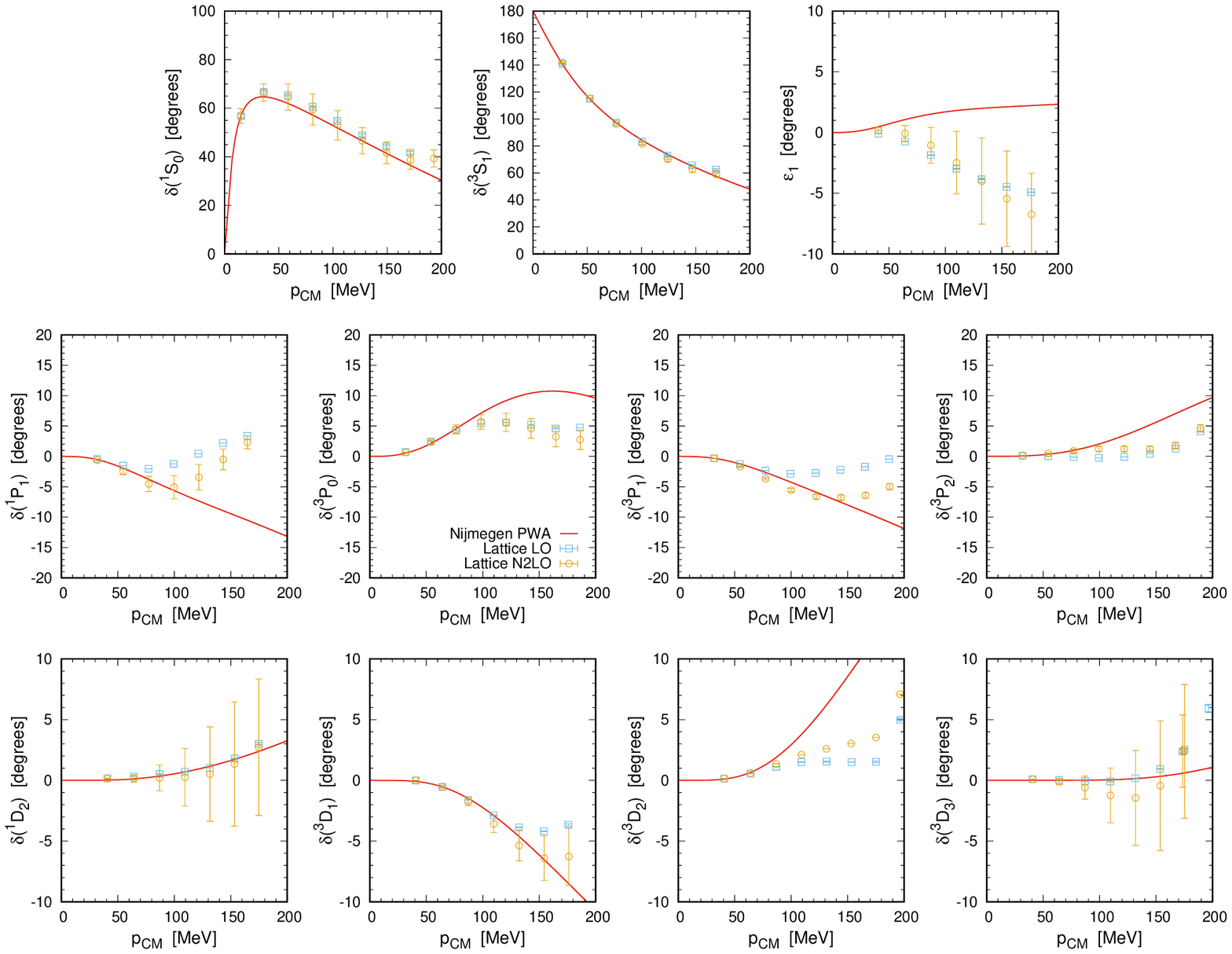}
\caption{LO (squares) and NNLO (circles) neutron-proton phase shifts and 
mixing angles for $a = 1.97$~fm. The NPWA is given by the solid line.
\label{Fig:Plot_pert_100}}
\end{center}
\end{figure*}

\begin{figure*}[hbt]
\begin{center}
\includegraphics[width=\textwidth]{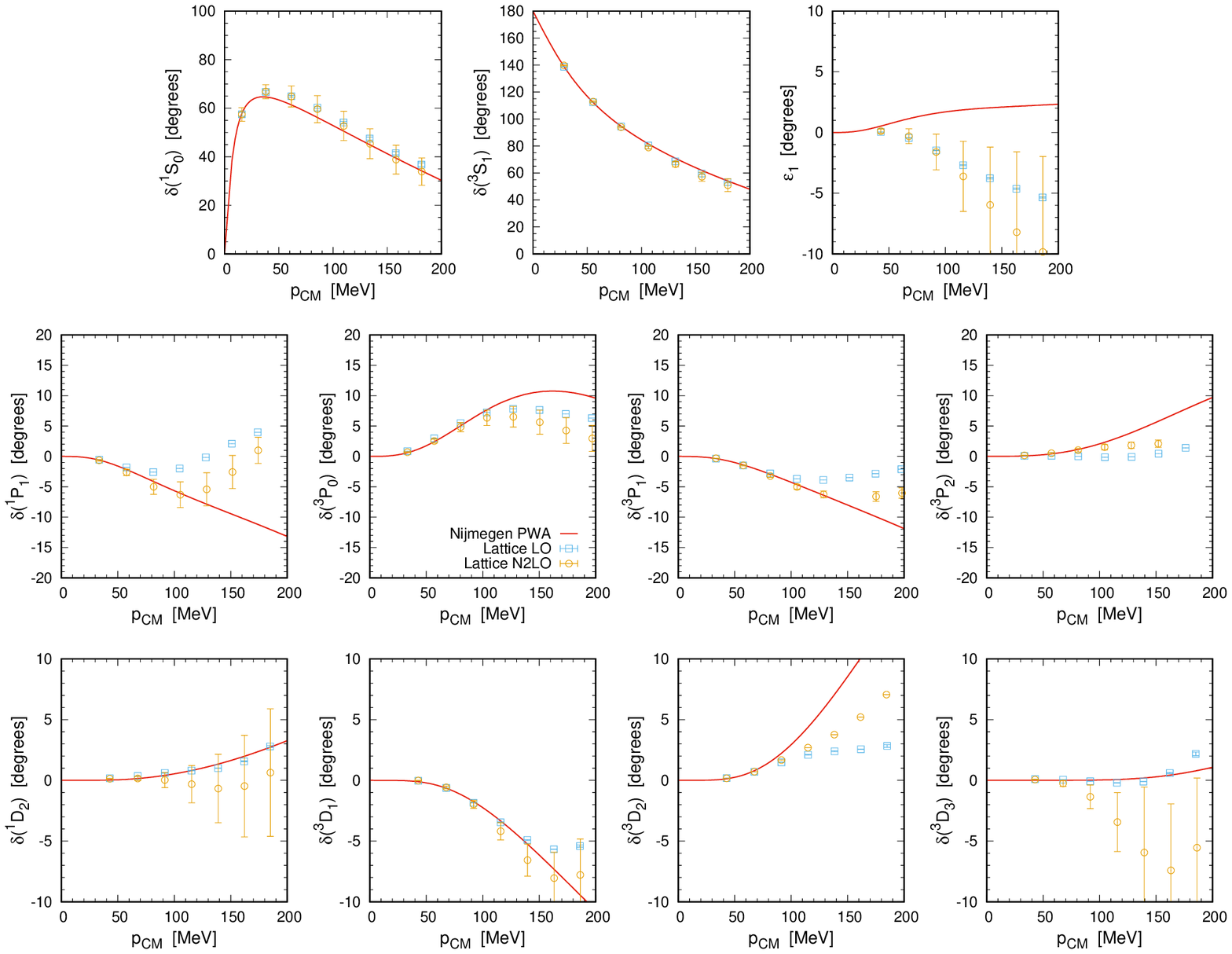}
\caption{LO (squares) and NNLO (circles) neutron-proton phase shifts and mixing angles for $a = 1.64$~fm. The NPWA is given by the solid line. \label{Fig:Plot_pert120}}
\end{center}
\end{figure*}

\begin{figure*}[hbt]
\begin{center}
\includegraphics[width=\textwidth]{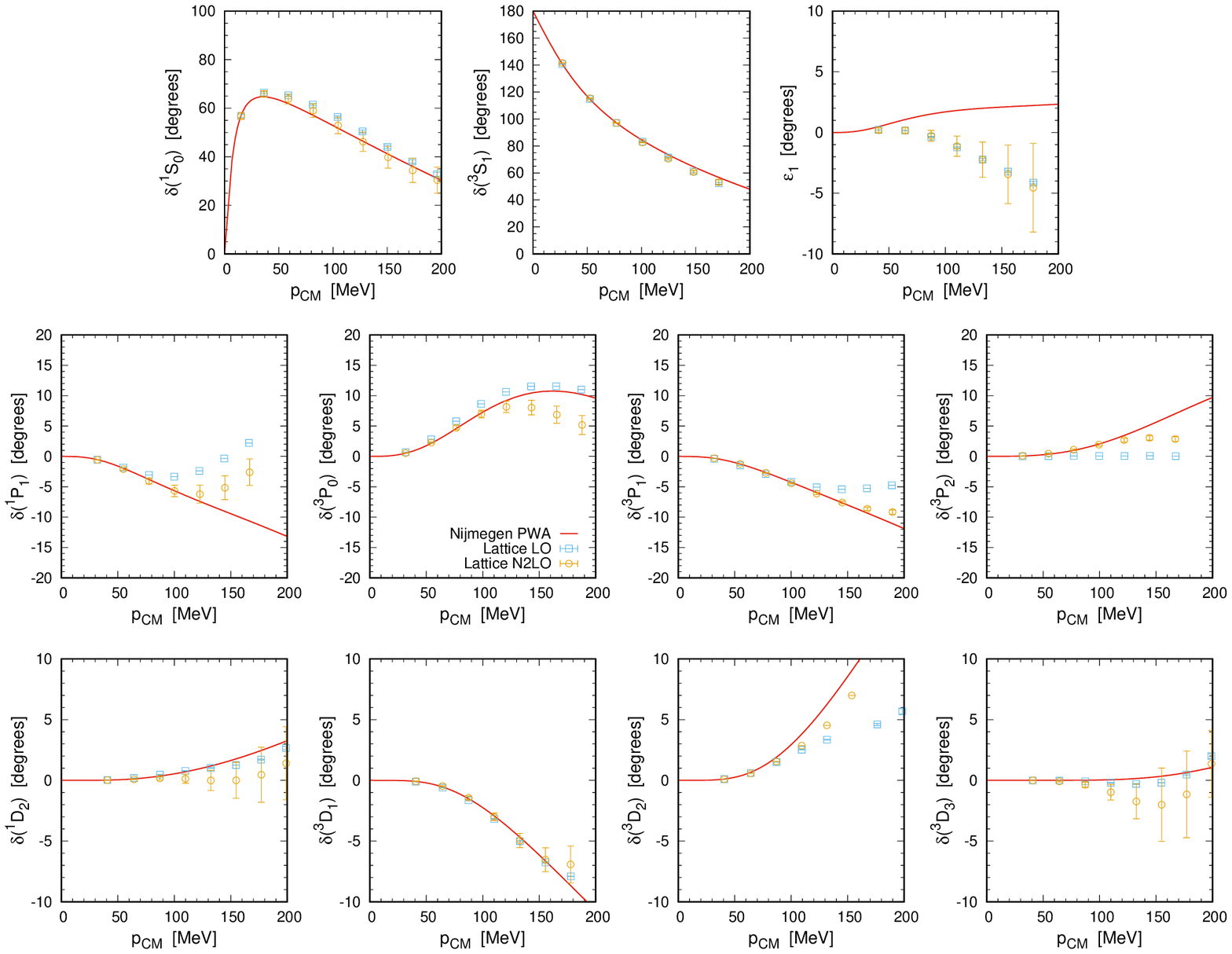}
\caption{LO (squares) and NNLO (circles) neutron-proton phase shifts and mixing angles for $a = 1.32$~fm. The NPWA is given by the solid line. \label{Fig:Plot_pert150}}
\end{center}
\end{figure*}

\begin{figure}[hbt]
\begin{center}
\includegraphics[width=0.48\textwidth]{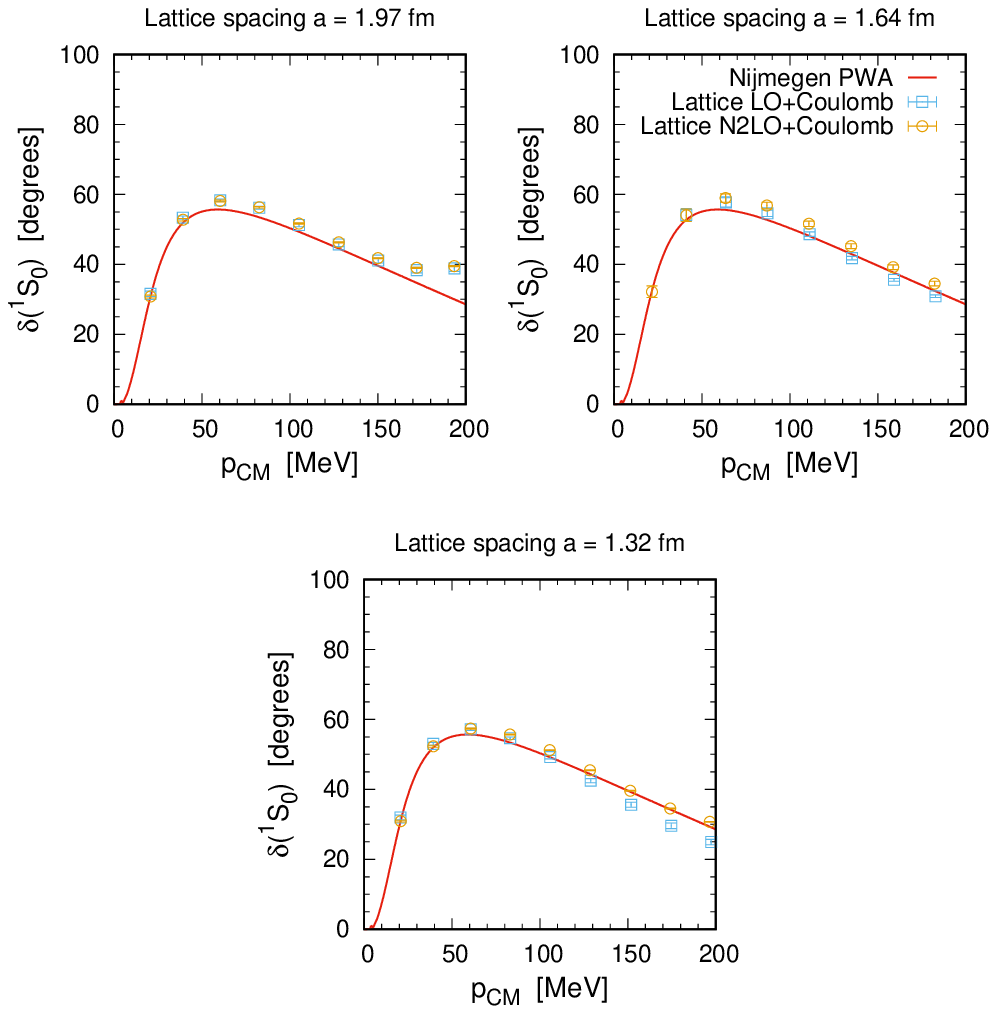}
\caption{LO (squares) and NNLO (circles) (both including Coulomb) proton-proton $^1S_0$ phase shift for 
$a = 1.97$~fm, $a = 1.64$~fm and $a = 1.32$~fm. The NPWA is given by the solid line. 
\label{Fig:Plot_pert_pp_1S0}}
\end{center}
\end{figure}

The results for np scattering can be found in Figs.~\ref{Fig:Plot_pert_100},\ref{Fig:Plot_pert120},\ref{Fig:Plot_pert150} while the pp scattering results are shown in Fig.~\ref{Fig:Plot_pert_pp_1S0}.
The corresponding LECs are summarized in Tab.~\ref{Tab:Perturbative_LECsDe_DX}.

\begin{table}[t]
\begin{center}
\caption{Summary of fit results with perturbatively 
improved OPE (in units of $a$) for the perturbative NLO+NNLO analysis at $a=1.97$~fm.
All LECs are given in lattice units.} 
\label{Tab:Perturbative_LECsDe_DX}
\smallskip
\begin{tabular*}{0.475\textwidth}{@{\extracolsep{\fill}}lrrr}
\hline\hline
\noalign{\smallskip}
    & $a=1.97$~fm& $a=1.64$~fm & $a=1.32$~fm  
\smallskip \\
 \hline
  $C_{^1S_0}^{}$         &   $-0.421(2)$    & $-0.370(4)$    &    $-0.289(3)$ \\
  $C_{^3S_1}^{}$         &   $-0.603(2)$    & $-0.549(3)$    &    $-0.424(3)$ \\
  \hline
  $\Delta C$             &  $-0.2(2)$           & $0.4(3)$   &    $-0.5(2)$   \\
  $\Delta C_{I^2}^{}$    &  $-0.0(1)$           & $-0.0(1)$  &    $-0.08(9)$  \\
  $C_{q^2}^{}$           &  $-0.05(6)$          & $0.10(8)$  &    $0.45(6)$   \\
  $C_{I^2, q^2}^{}$      &  $-0.06(3)$          & $0.13(4)$  &    $0.27(4)$   \\
  $C_{S^2, q^2}^{}$      &  $-0.00(7)$          & $-0.0(1)$  &    $0.00(5)$   \\
  $C_{S^2, I^2, q^2}^{}$ &  $-0.01(5)$          & $0.0(1) $  &    $-0.10(3)$  \\
  $C_{(q\cdot S)^2}^{}$  &   $0.00(8)$          & $-0.0(0)$  &    $0.05(3)$   \\
  $C_{I^2, (q\cdot S)^2}^{}$ &   $0.02(8)$      & $0.2(1) $  &    $0.20(3)$   \\
  $C_{(q \times S)\cdot k}^{I=1}$ &  $0.033(6)$ & $0.04(1)$  &    $0.09(1)$   \\
  $C_{SO(3)}$             &  $0.1(1)$           & $1(1)   $  &    $-0.5(1)$   \\
\hline
  $C_{nn}$               & $0.01(4)$            & $0.02(5)$  &    $0.04(5)$   \\
  $C_{pp}$               & $0.003(1)$          & $0.003(1)$&    $0.008(3)$ \\
  \hline 
  $E_{\rm LO}$~[MeV]         & $ -2.20(4)$          & $-2.42(8)$ &  $-2.6(1)$\\ 
 \hline\hline                          
\end{tabular*}
\end{center}
\end{table}

When we compare the LO results for the various lattice spacings, we see that the $^1S_0$ phase shift 
is too strong already at 70~MeV while the description of the $^3S_1$ phase shift is quite accurate 
even beyond the fit range of 100~MeV, but the best description is for $a= 1.64$~fm instead of 
$a= 1.32$~fm. The reason is that the smearing constant $b$ is fixed instead of a fit parameter which results also in a fixed shape of the $^3S_1$ phase shift. 
The calculated P-wave phase shifts are in agreement with the PWA phase shifts roughly up to 
80~MeV for $a=1.97$~fm and the description improves with smaller lattice spacing. As the only 
influence is from the OPE, this does not come as a surprise as the simplified description of the 
OPE numerator approaches more and more the exact one with smaller lattice spacings.
Also the D-wave description at LO improves significantly, e.g. the $^3D_1$ channel description 
is quite fine up to 80~MeV for the coarse lattice while it is quite good up to 170~MeV for the 
fine lattice. At N2LO the general description improves as the fit range is extended up to 
150~MeV and also the P-waves as well as the deuteron binding energy is included. This 
improvement can be seen particularly in the $^1S_0$ channel where the phase shift moves 
closer to the PWA analysis and $^1P_1$ where the agreement range is extended by 40~MeV. 
Comparing the different lattice spacings, one sees again a clear improvement particularly 
for the P-waves which are now described up to the fit range of $150$~MeV. For D-waves 
of the smallest lattice spacing we still have some small deviations at least for the $^1D_2$ 
channel as well as the $^3D_3$ channel which would be fixed by the inclusion of N3LO corrections.

Having a closer look at the pp $^1S_0$ phase shift one sees that the phase shift is too large for $a=1.97$~fm particularly for high energies. While the phase shift becomes smaller and finally too small with finer lattices at LO, the N2LO phase shifts is getting more and more close to the NPWA phase shift in the whole momentum region from 0 to 200~MeV.

\section{Three-body sector}\label{sec:threebody}

In the three-body sector we only have to consider the triton. Its experimental binding energy 
is given by $E_{^3\rm H}=-8.4820(1)$~MeV.

\subsection{Theoretical framework}
Although we work in the three-body sector, we are still able to do an exact calculation 
using the Lanczos method. Therefore we extend the former analysis of the 2N sector to 
the triton, which means that we cannot work in radial coordinates anymore but we 
calculate the spectrum in three dimensions where we also include three-body potentials, 
which were reviewed in \cite{Epelbaum:2009pd}. They consist of a three-body contact interaction, 
a one- and a two-pion exchange interaction. These various terms read:
\begin{align}
V^{\rm 3N}_{\rm contact} &= D_{\rm contact}^{\rm 3N}\sum_{\vec{n}}\colon\rho
\left(\vec{n}\right)\rho\left(\vec{n}\right)\rho\left(\vec{n}\right)\colon~,\\
\begin{split}
V^{\rm 3N}_{\rm OPE} &= D_{\rm OPE}^{\rm 3N}\sum_{\vec{n}_{1},S_1,\vec{n}_{2},S_2,I} \left[ 
G_{S_1,S_2}\left(\vec{n}_1-\vec{n}_2\right)\right.\\
&\left.\times \colon\rho_{S_1,I}\left(\vec{n}_1 \right) \rho_{S_2,I}\left(\vec{n}_2 \right) 
\rho \left(\vec{n}_2 \right)\colon \right]~,
\end{split}\\
V^{\rm 3N}_{\rm TPE} &= V^{\rm 3N}_{\rm TPE,m^2}+V^{3N}_{\rm TPE,p^2}+V^{3N}_{TPE,xx}.
\end{align}

The latter equation describing the TPEP among three particles can be split up into three parts 
which read
\begin{align}
\begin{split}
V^{\rm 3N}_{\rm TPE,q^2}&= D_{q^2}^{\rm 3N}\sum_{\substack{\vec{n}_1,S_1, \vec{n}_2,S_2,\\ \vec{n}_3,S_3,I}} 
\left[ G_{S_1,S_3}\left(\vec{n}_1-\vec{n}_3\right)  \right.\\
 \times G_{S_2,S_3}& \left.\left(\vec{n}_2-\vec{n}_3\right) \colon\rho_{S_1,I}\left(\vec{n}_1 \right) \rho_{S_2,I}\left(\vec{n}_2 \right) \rho \left(\vec{n}_3 \right)\colon \right]~,
\end{split}\\
\begin{split}
V^{\rm 3N}_{\rm TPE,m^2}&=D_{m^2}^{\rm 3N}\sum_{\substack{\vec{n}_1,S_1, \vec{n}_2,S_2, \vec{n}_3,I}} 
\left[ G_{S_1}\left(\vec{n}_1-\vec{n}_3\right)  \right.\\
\times G_{S_2}&\left.\left(\vec{n}_2-\vec{n}_3\right) \colon\rho_{S_1,I}\left(\vec{n}_1 \right) \rho_{S_2,I}\left(\vec{n}_2 \right) \rho \left(\vec{n}_3 \right)\colon \right]~,
\end{split}\\
\begin{split}
V^{\rm 3N}_{\rm TPE,xx}&=D_{xx}^{\rm 3N}\sum_{\substack{\vec{n_3},\tilde{S}_1,\tilde{S}_2,\tilde{S}_3, I_1,I_2, \\I_3,\vec{n}_2,S_2,\vec{n}_3,S_3}} \left[ G_{S_1,\tilde{S}_1}\left(\vec{n}_1-\vec{n}_3\right) \right. \\
&\times G_{S_2,\tilde{S}_2}\left(\vec{n}_2-\vec{n}_3\right) \epsilon_{\tilde{S}_1,\tilde{S}_2,\tilde{S}_3} \epsilon_{I_1,I_2,I_3} \\
&\left.\times\colon\rho_{S_1,I_1}\left(\vec{n}_1\right)\rho_{S_2,I_2}\left(\vec{n}_2\right)\rho_{\tilde{S}_3,I_3}\left(\vec{n}_3\right)\colon \right]~,
\end{split}
\end{align}
where the coefficients are
\begin{align}
&D_{\rm contact}^{\rm 3N}= \frac{-3c_E}{F_\pi^4\Lambda}, 
\qquad D_{\rm OPE}^{\rm 3N}= \frac{c_D}{4 F_\pi^3\Lambda}\frac{g_A}{2F_\pi },\nonumber\\
&D_{q^2}^{\rm 3N} = \frac{c_3}{F_\pi^2}\frac{g_A^2}{4F_\pi^2 },
\qquad D_{m^2}^{\rm 3N} = \frac{-2c_1}{F_\pi^2}\frac{M_\pi^2 g_A^2}{4F_\pi^2 }, \nonumber\\
&D_{xx}^{\rm 3N} = \frac{c_4}{2F_\pi^2}\frac{g_A^2}{4F_\pi^2 }, 
\end{align}
with $\Lambda=700$~MeV as the reference scale. There are two new dimensionless parameters 
$c_D$ and $c_E$ which must be determined using at least two three-body observables. While 
we use the well-measured triton binding energy as one parameter, it was summarized in \cite{Epelbaum:2009pd} that $c_D$ and $c_E$ could be disentangled by additionaly including of 
nucleon-deuteron scattering, triton beta decays or some other observable in the analysis. 
As this is beyond the scope of this work, we keep the correlation between $c_D$ and $c_E$ and 
we fix $c_D=-0.79$ as it was shown to be of $\mathcal{O}(1)$. Hence, we use $c_E$ as the only 
fit parameter and fix it with the triton binding energy of $E_B^{^3H}=-8.4820(1)$~MeV. In 
the subsequent part of this paper we also have a look on systematic errors due to this 
particular choice of $c_D$. Of course, there are better ways of fixing $c_D$ by now, but for the
sake of consistency we have to use the same method that was employed in earlier NLEFT
calculations

As we are interested in the binding energy of the system, we have to calculate the ground 
state of the system at large enough volume or do a finite volume extrapolation for a 
three-particle system. Using a box volume of $V \approx (10 \cdot 1.97~\mathrm{fm})^3 
\approx (12 \cdot 1.64~\mathrm{fm})^3 \approx (20~\mathrm{fm})^3$ for the two coarsest 
lattices is enough for neglecting the finite volume effects and it is still calculable with 
in a reasonable amount of computational ressources. Unfortunately, this volume is not 
computable with the given resources anymore for a lattice spacing of $a=1.32$~fm as the 
problem scales $\propto L^6$. Finite volume binding energy corrections for three particles were calculated in the unitary limit as well as the shallow binding of one particle to a deeply bound dimer in Refs.~\cite{Meissner:2014dea,Konig:2017krd,Hammer:2017uqm}. While the triton is a system between these two limits, the numerical difference between the two calculations is negligible once the volume is chosen large enough. Hence, we do a finite volume extrapolation using the LO formula in the unitary limit given by 
\begin{equation}
E^{\rm 3N}\left(L\right)=E^{\rm 3N}_\infty+\mathcal{A}\frac{\exp\left(\frac{2\kappa L}{\sqrt{3}}\right)}
{\left(\kappa L\right)^{\frac{3}{2}}}
\end{equation}
with $\kappa=\sqrt{-mE^{\rm 3N}_\infty}$ and using all data points for $L=10$ and larger as 
for smaller lattices the NLO contributions of the finite volume corrections become significant. 
Afterwards, we fit each perturbative higher-order operator $\braket{\mathcal{O}_i}$ 
according to 
\begin{equation}
\braket{\mathcal{O}_i}\left(L\right)=\braket{\mathcal{O}_i}_\infty
+\mathcal{A}_i\frac{\exp\left(\frac{2\kappa L}{\sqrt{3}}\right)}{\left(\kappa L\right)^{\frac{3}{2}}}.
\end{equation}

\subsection{Results}

\begin{figure}[hbt]
\begin{center}
\includegraphics[width=0.45\textwidth]{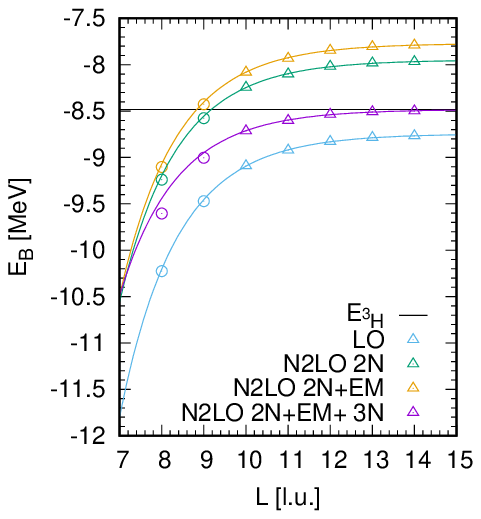}
\caption{Finite volume extrapolation for the triton and a lattice spacing of $a = 1.32$~fm. 
The triangles are included in the fit while the circles are not due to higher order 
finite volume effects. \label{Fig:triton_pert150_fV}}
\end{center}
\end{figure}
The extrapolation is plotted in Fig.~\ref{Fig:triton_pert150_fV} where one can see excellent 
agreement with the data points. The fit quality also makes it unnecessary to include higher 
order corrections use other methods like twisted mass boundary conditions to further pin down 
the infinite volume binding energy \cite{Korber:2015rce}. The results for the various 
lattice spacings are summarized in Tab.~\ref{Tab:triton_pertall}. 
\begin{table}[t]
\begin{center}
\caption{Triton binding energy predictions at LO, N2LO, N2LO+EM, N2LO+EM+3N and the fit 
parameter $c_D$. The energy errors in brackets are due to the uncertainties of the LECs.} 
\label{Tab:triton_pertall}
\smallskip
\begin{tabular*}{0.475\textwidth}{@{\extracolsep{\fill}}lrrr}
\hline\hline
\noalign{\smallskip}
    & $a=1.97$~fm& $a=1.64$~fm & $a=1.32$~fm  
\smallskip \\
 \hline
  $E_{\rm LO}$~[MeV]         &   $-7.80$    & $-8.29$    &    $-8.74$ \\
  $E_{\rm N2LO}$~[MeV]          &   $-7.846(4)$ & $-8.11(2)$ &    $-7.95(2)$ \\
  $E_{\rm N2LO}^{\rm +EM}$~[MeV]    &   $-7.68(2) $ & $-7.91(3)$ &    $-7.77(2)$   \\
  $E_{\rm N2LO}^{\rm +EM+3N}$\footnote{At this order, the triton binding energy is a fit parameter.}~[MeV] &   $-8.48(3)$ & $-8.48(3)$ &    $-8.48(2)$ \\
  \hline
  $c_E$      &   $0.5309(2)$     &  $0.3854(3)$       &    $1.0386(5)$ \\
 \hline\hline
\end{tabular*}
\end{center}
\end{table}
Focusing on the LO, one sees an underbinding at LO of only $-7.80$~MeV for $a=1.97$~fm, an almost 
perfect binding energy of $-8.29$~MeV for $a=1.64$~fm and an overbinding of $-8.74$~MeV for the smallest 
lattice spacing of $a=1.32$~fm. By comparing these results with the neutron-proton phase shifts, one 
can attribute this mainly to the $^3P_0$ phase shift where one has a strong shift from its 
underestimation of it at the coarse lattice spacing to its overestimation at the fine one. At N2LO, 
the binding energy varies around 8~MeV. In particular, the triton becomes less bound as $a$ is decreased
from $a=1.64$~fm to $a=1.32$~fm even though the $^3P_0$ prediction is stronger. The reason is that the 
difference between the $^3P_0$ phase shifts is relatively small and the $^1S_0$ as well as the $^3P_1$ 
phase shifts become smaller and finally have a larger effect on the three particle binding energy. 
The fit value for $c_E$ is of natural size and its pattern is consistent
with the missing attraction at N2LO+EM.

\section{Four-body sector}\label{sec:fourbody}
\subsection{Theoretical framework}
In the four-body system, we do not have any new operator as our system should be describable 
by the 2NFs and 3NFs only. As the four-body system scales with $L^9$, an exact calculation at sufficient large lattices is not practical anymore, and hence we have to use Monte Carlo methods. 
More precisely, we use auxiliary field Monte Carlo with the hybrid Monte Carlo 
algorithm \cite{Lee:2008fa}.
In the following we will define the LO auxiliary field transfer matrix which we will 
minimize afterwards. All other contributions are calculated perturbatively.
For an increased convergence we prepare our trial states using a SU(4) symmetric Hamiltonian
\begin{equation}
H_0=H_{\rm free}+\frac{1}{2}C_0\sum_{\vec{n}_1,\vec{n}_2} f\left(\vec{n}_1-\vec{n}_2\right)
\rho\left(\vec{n}_1\right)\rho\left(\vec{n}_2\right),
\end{equation}
with $f(\vec{n}_1-\vec{n}_2)$ a Gaussian smearing function. This operator is used to efficiently 
create trial states which are close to realistic nuclei,
\begin{equation}
\ket{\psi_{^4\text{He}}}=\exp\left(-t_0H_0\right)\ket{\psi_{0}},
\end{equation}
with $\ket{\psi_{0}}$ the antisymmetrized free-particle solution for ${}^4\text{He}$ in a finite volume. The correlation function is defined as 
\begin{equation}
Z_{^4\text{He}}\left(t\right)=\braket{\psi_{^4\text{He}}\mid\exp(-t H_{\rm LO})\mid\psi_{^4\text{He}}}~,
\label{eq:psiout}
\end{equation}
where $H_{\rm LO} = H_{\rm free}+H_{\rm LO,contact}+H_{\rm OPE}$ is the full LO Hamiltonian according 
to Eqs.~(\ref{eq:Hfree},\ref{eq:HLOcontact},\ref{eq:HOPE}), and $\ket{\psi_{^4\text{He}}}$ is the 
antisymmetrized wave function of the nucleons given by Eq.~(\ref{eq:psiout}). The above-mentioned expression can be calculated using auxiliary field Monte Carlo methods for 
different time steps and the corresponding energy is given by 
\begin{equation}
E_{\rm LO}\left(t\right)=-\frac{d\log Z_{^4\text{He}}\left(t\right)}{dt}.
\end{equation}
The correlation function for any perturbative operator $\mathcal{O}$ is defined by 
\begin{equation}
Z_\mathcal{O}\left(t\right)=\braket{\psi_{^4\text{He}}\mid \exp\left(\frac{-tH}{2}\right) 
\mathcal{O}\exp\left(\frac{-tH}{2}\right)\mid\psi_{^4\text{He}}},
\end{equation}
and their expectation value is given by the ratio
\begin{equation}
\braket{\mathcal{O}}\left(t\right)=\frac{Z_\mathcal{O}\left(t\right)}{Z_{^4\text{He}}(t)}.
\end{equation}
The ground state energy is calculated by performing the Euclidean time extrapolation to the infinity. Therefore we fit LO, additional 2N N2LO, additional 2N electromagnetic and additional 2N N2LO contribution separately with one or two exponential decay functions depending on the contribution and sum them up finally.
\begin{equation}
\begin{split}
E_\mathcal{O}(t)=E_{0,\mathcal{O}}&+c_1\exp\left(-\Delta E_{1,\mathcal{O}} t\right)\\
\left[\right.&+\left. c_2\exp\left(-\Delta E_{2,\mathcal{O}}t\right)\right] .
\end{split}
\end{equation}
The necessity of two or even more exponentials for the extrapolation of perturbative operators 
was already shown in \cite{Lahde:2013uqa}, where an analysis with particular emphasis on 
the infinite time extrapolation was done.
In the following we do a benchmark calculation for $L = 4$ and $a=1.97$~fm which we can compare 
with an exact Lanczos calculation. Then we do the calculation again for $L = 6$ and $a=1.97$~fm, 
$L = 7$ and $a = 1.64$~fm and $L = 9$ and $a =1.32$~fm. Then the physical box length is 
between 11~fm and 12~fm and it is large enough that finite volume errors will be within 
truncation errors due to chiral expansion and uncertainties in the respective low-energy 
coupling constants (LECs).

\subsection{Results}
\begin{figure}[ht!]
\begin{center}
\includegraphics[width=0.48\textwidth]{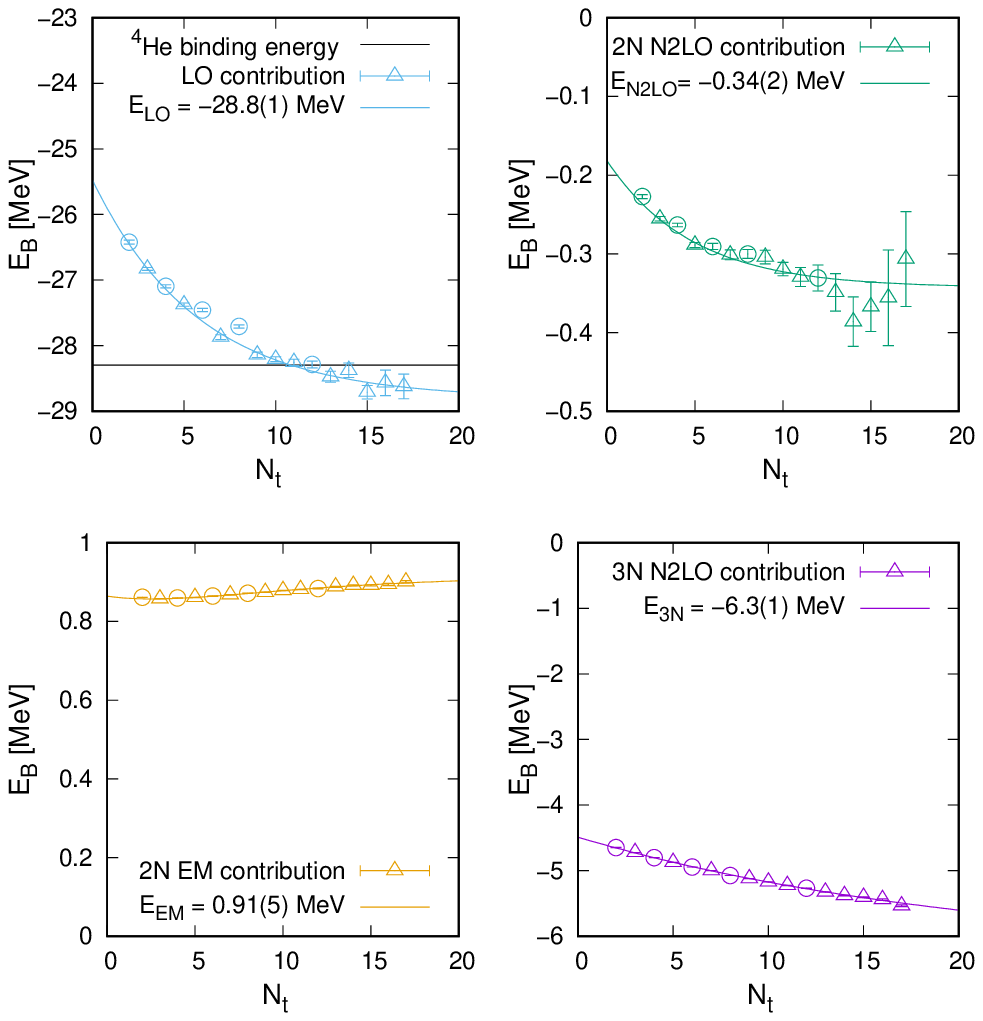}
\caption{$^4$He time extrapolation for lattice spacing of $a = 1.97$~fm. The triangles are included in the fit while the circles are excluded due to bad statistics. \label{Fig:He4_MC_100}}
\end{center}
\end{figure}

\begin{figure}[ht!]
\begin{center}
\includegraphics[width=0.48\textwidth]{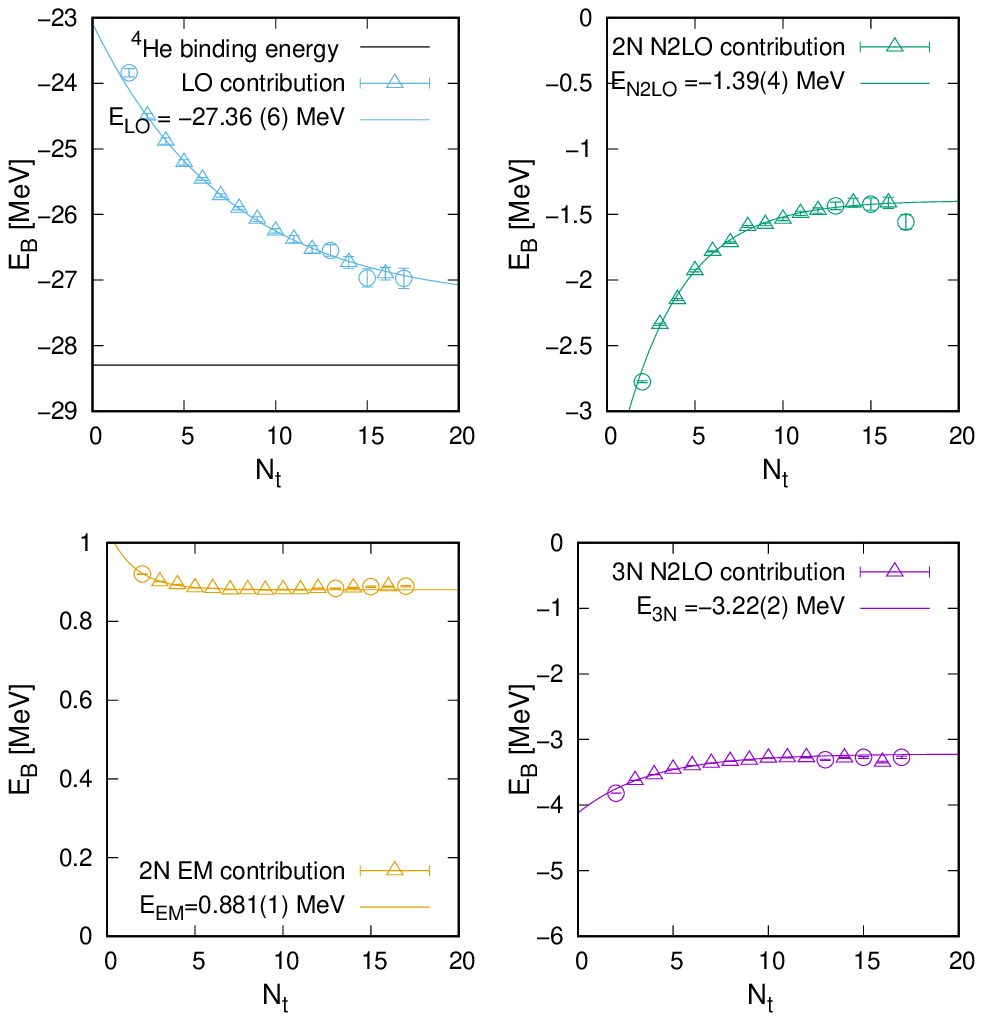}
\caption{$^4$He time extrapolation for lattice spacing of $a = 1.64$~fm. The triangles are included in the fit while the circles are excluded due to bad statistics. \label{Fig:He4_MC_120}}
\end{center}
\end{figure}

\begin{figure}[ht!]
\begin{center}
\includegraphics[width=0.48\textwidth]{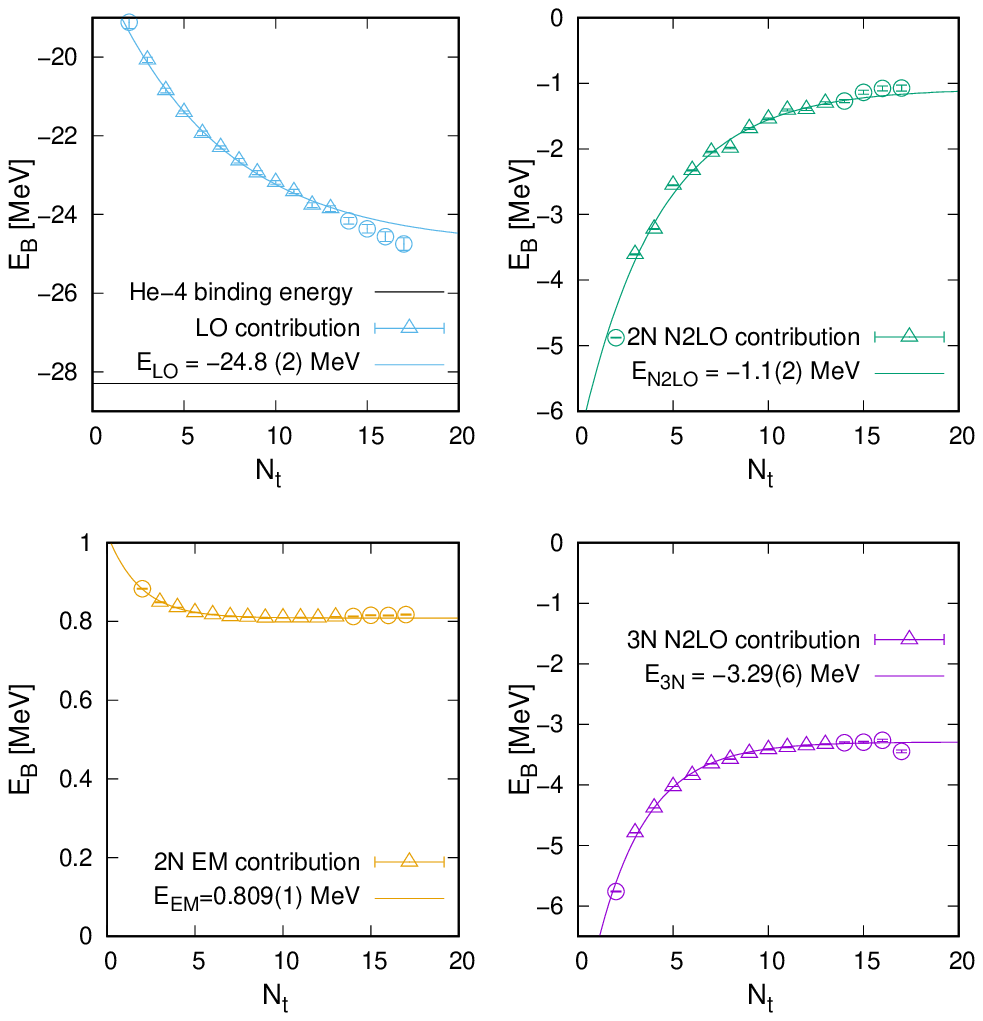}
\caption{$^4$He time extrapolation for lattice spacing of $a = 1.32$~fm. The triangles are included in the fit while the circles are excluded due to bad statistics. \label{Fig:He4_MC_150}}
\end{center}
\end{figure}

First of all, we start with the benchmark calculation. The results are shown in 
Tab.~\ref{tab:He4_100_L4}. One can see very good agreement particularly for the LO result 
which is around one per mille relative error. The difference for the perturbative corrections 
is larger but still below 10~\% which is finally within the error bars of the infinite 
time LO extrapolation. Even though the accuracy will go down with larger volumes due to the 
sign problem we do expect trustable results within our estimated errors. The finite time 
extrapolation order by order for the three lattice spaces are shown in 
Figs.~\ref{Fig:He4_MC_100}, \ref{Fig:He4_MC_120} and \ref{Fig:He4_MC_150} while the 
summed binding energy predictions are shown in Tab.~\ref{Tab:helium_pertall}. First,
one can see very good time extrapolation order by order for all three lattice spacings. 
While the statistical errors at LO are below 1\%, the perturbative relative errors are 
around 3\% except for the N2LO contribution 
for $a=1.32$~fm where the error is much larger. This is caused by 
relatively bad statistics of the data points due to the very large lattice used. As the 
higher-order contributions are quite small, their error due to statistical uncertainties 
as well as uncertainties in the LECs are dominated by the LO statistical uncertainties as 
shown in Tab.~\ref{Tab:helium_pertall}. As mentioned in Sec.~\ref{sec:threebody}, there is 
some ambiguity in the determination of $c_D$ and $c_E$ where $c_D$ is of $\mathcal{O}(1)$. 
We therefore  fitted $c_E$ for different values of $c_D$, namely $-2 \leq c_D \leq 2$. The 
difference in the $^4$He binding energy comes out very small, $\Delta E^{^4\mathrm{He}}_{c_D-c_E}\approx 0.2$~MeV. 
This systematic error is 
compatible with the errors caused by statistics as well as the uncertainties of the NLO and N2LO LECs.
\begin{table}[t]
\begin{center}
\caption{$^4$He benchmark calculation for $a=1.97$~fm and $L = 4$. The first bracket of Monte Carlo error are statistical ones, while the latter ones and the Lanczos ones are errors due to uncertainties of the LECs. } 
\label{tab:He4_100_L4}
\smallskip
\begin{tabular*}{0.475\textwidth}{@{\extracolsep{\fill}}lrr}
\hline\hline
\noalign{\smallskip}
    & Monte Carlo& Lanczos 
\smallskip \\
 \hline
  $E_{\rm LO}$~[MeV]           &   $-30.32(2)(1)$    & $-30.34$  \\
  $\Delta E_{\rm N2LO}$~[MeV]  &   $ 0.511(9)(10)$   & $ 0.52(2)$  \\
  $\Delta E_{\rm EM}$~[MeV]    &   $0.86(4)(2)$     & $0.91(3)$ \\
  $\Delta E_{\rm 3NNLO}$~[MeV] &   $-5.1223(1128)(5)$& $-5.0278(5)$  \\
 \hline\hline
\end{tabular*}
\end{center}
\end{table}

\begin{table}[t]
\begin{center}
\caption{$^4$He binding energy prediction at LO, N2LO, N2LO+EM, N2LO+EM+3N. 
The first brackets give the statistical error while the latter ones give 
the errors due to uncerainties of the LECs. All binding energies are given in MeV.} 
\label{Tab:helium_pertall}
\smallskip
\begin{tabular*}{0.48\textwidth}{@{\extracolsep{\fill}}lrrr}
\hline\hline
\noalign{\smallskip}
    & $a=1.97$~fm& $a=1.64$~fm & $a=1.32$~fm  
\smallskip \\
 \hline
  $E_{\rm LO}$            &   $-28.81(11)$ & $-27.36(6)$ & $-24.81(19)$ \\
  $E_{\rm N2LO}$          &   $-29.15(11)(3)$ & $-28.75(7)(5)$ & $-25.89(27)(3)$ \\
  $E_{\rm N2LO}^{\rm +EM}$    &   $-28.23(12)(3)$ & $-27.87(7)(6)$ & $-25.08(27)(3)$ \\
  $E_{\rm N2LO}^{\rm +EM+3N}$ &   $-34.55(18)(3)$ & $-31.09(7)(6)$ & $-28.37(28)(3)$ \\
 \hline\hline
\end{tabular*}
\end{center}
\end{table}

\section{The Tjon band}\label{sec:Tjon}
\subsection{Theoretical framework}

The correlation between the $^3$H and $^4$He binding energies was first observed by 
Tjon~\cite{Tjon:1975sme} for a large class of 2N potentials of different accuracy.
This was later dubbed the Tjon line.
It was shown in Refs.~\cite{Nogga:2000uu, Epelbaum:2002vt, Nogga:2004ab} that this correlation 
still holds in the case of modern, accurate semi-phenomenological potentials as well as 
nuclear effectice field theory. In Ref.~\cite{Platter:2004zs} this correlation was studied 
in the framework of pionless effective field theory, where the only input parameters are 
the singlet and triplet neutron-proton scattering lengths as well as the deuteron binding
energy $E_d$. 
In this study, it was also possible to give a range for the correlation by calculating it 
either with $a_{^1S_0}$ and $a_{^3S_1}$ scattering lengths as input parameters or with $a_{^1S_0}$ 
and $E_d$ as input parameters. In this way, the so-called Tjon band is generated.
In Fig.~\ref{Fig:Tjonline} the upper bound is due to the first fit while the lower bound is 
due to the latter one. A similar analysis using resonating group techniques in the framework 
of pionless EFT was done in Ref.~\cite{Kirscher:2009aj}. However, in NLEFT a general overbinding was 
observed \cite{Epelbaum:2009pd,Lahde:2013uqa} in the case of a very coarse lattice of $a=1.97$~fm. 
This overbinding was systematically absorbed in an effective four-body contact interaction 
which was fitted to the binding energy of the alpha-cluster nucleus $^{24}$Mg. It was argued 
that this overbinding is a lattice arfefact which is caused by an implicit 4NF due to the 
superposition of four particles at the same space point. In general, this contribution is 
negligible in the continuum, but due to the binning of the wave function over the lattice 
point volume, this contribution may become unphysically large and contribute to very deep 
bound states. This was shown explicitly in two dimensions in Ref.~\cite{Lee:2005nm} and it 
should vanish once the lattice spacing is small enough.

\subsection{Results}

\begin{figure}[ht!]
\begin{center}
\includegraphics[width=0.48\textwidth]{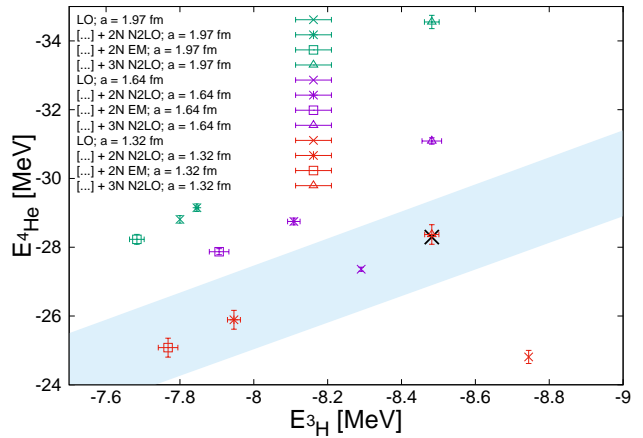}
\caption{$^3$H-$^4$He binding energy plot for various lattice spacings. The black dot is 
the physical point and 
the blue band is the Tjon band according to \cite{Platter:2004zs}. The $^4$He errors 
include statistical and LEC errors while the $^3$H errors only include 
LEC errors.\label{Fig:Tjonline}}
\end{center}
\end{figure}

The results for the binding energy of triton and $^4$He in the previous sections are 
combined and shown in Fig.~\ref{Fig:Tjonline}. For the standard coarse lattice spacing of 
$a=1.97$~fm already the LO is above the Tjon line as $^3$H is approximately $1$~MeV underbound 
or $^4$He is approximately 2.5~MeV overbound. The data points for 2N N2LO are close the LO data point as there is not very much difference in the np phase shift 
shown in Fig.~\ref{Fig:Plot_pert_100} as well. Including the 2N EM interaction results in a data point closer to the Tjon line but still around 2~MeV above. However, the inclusion of 3N N2LO contributions results in a very large overbinding of 
approximately 6~MeV. For the next lattice spacing 
of $a=1.64$~fm the results already become better as the LO data point is already on the 
Tjon band and the N2LO/ N2LO+EM correction is closer to the band as well but still above. After the inclusion of the 3N forces, the overbinding of $^4$He is 
only 2.5~MeV for $a=1.64$~fm. In the last case of the finest lattice spacing of $a=1.32$~fm, the LO triton 
binding energy is approximately $0.3$~MeV too strong while the $^4$He binding energy is 
roughly $4$~MeV too small. The respective $^4$He-$^3$H data point is now below the Tjon 
line which does not come as a suprise. The reason is that for a good description within the 
Tjon band it is necessary to have a very good description of the $^1S_0$, $^3S_1$ phase shifts and the deuteron 
binding energy. By comparing the LO deuteron binding energies summarized in 
Tab.~\ref{Tab:Perturbative_LECsDe_DX} one can see an appearing overbinding for 
smaller lattice spacings. Such a overbinding should lead to a decrease of the Tjon 
band towards the measured $^3$H-$^4$He energy. Once higher orders are included the 
binding energy is fixed and the results are within the Tjon band around $-7.95$~Mev 
for the triton and $-25$~MeV for $^4$He. The electromagnetic contributions shift the data point but it is still almost in the center of the Tjon band. After including the 3NF, the triton energy is at 
the physical point and the $^4$He energy is $-28.37(28)$~MeV within the Tjon band. This means that 
all lattice artefacts are systematically removed and one can reproduce the correlation between 
the three- and a four-body system. As the physical point is already within the error bands, one would need more statistical improvement as this is the main error source. Then one can observe the influence of other remaining possible issues like more accurate N3LO np data or the ambiguity in the determination of  $c_D$ and $c_E$.
 
\section{Conclusion}\label{sec:conclusions}

In this paper we have analysed the Tjon band in the framework of NLEFT. We studied the two-,
 three-, and four-body sector for lattice spacings from $a=1.97$~fm to $a=1.32$~fm up to 
N2LO and including subleading two-pion-exchange contributions as well as the electromagnetic 
interaction and the leading 3NFs. There is a general convergence of the phase shifts in the 
two-body sector by including higher orders as well as shifting to smaller lattice spacings. 
In the three-body sector we found almost similar results at N2LO for all three-cases. 
The reason for this is that the three-body bound state is not sensitive to all np-phase 
shifts in the same way and even though the description of the phase shifts became better in general, 
some particular phase shifts do not improve leading to the 
the general underbinding of the system. In the $^4$He system we observed a strong 
overbinding of about 6~MeV due to lattice effects for the coarse lattice which becomes 
smaller with decreasing lattice spacings and vanishes finally. By comparing the triton and 
$^4$He binding energies for each lattice spacing and each order, one can see a 
convergence towards the Tjon line with smaller lattice spacing after including 
N2LO-forces, N2LO+EM-forces and N2LO+EM+3N forces, respectively. Finally the Tjon line 
is hit and confirming the conjecture that light (and medium mass) nuclei can be described 
by 2NFs and 3NFs only. The inclusion of 3N forces give a helium-4 binding energy prediction of 
$E_B=-28.37(28)(4)$~MeV which is consistent with the experimental value, $E_{^4_{\text{He}}}^{\text{exp}}=-28.30$~MeV.
Even though the deviation from the Tjon band vanishes for small lattice spacings, further 
investigation of these implicit multi-particle interactions is necessary as these small 
lattice spacings require very expensive computational ressources due to the increased number of 
nodes necessary for a reasonable volume. Further improvements on the results discussed
here can be obtained by improved statistics particularly for the smallest lattice spacing, by more more accurate N3LO np and pp phase shifts and also from
more detailed studies of the discretization effects arising from variations of the 
ratio $a^2/a_t$ that was kept fixed here. Finally, a reassesment of the determination
of the 3NFs LECs $c_D$ and $c_E$ would be useful.

\vfill

\section*{Acknowledgments}

This work was supported in part by the DFG and NSFC through funds provided to the
Sino-German CRC 110 ``Symmetries and the Emergence of
Structure in QCD'' (NSFC Grant No. 11621131001, DFG Grant No. TRR110), by the VolkswagenStiftung (Grant No. 93562), by the BMBF (contracts No. 05P2015 - NUSTAR R\&D) and by the CAS President's International Fellowship
Initiative (PIFI) (Grant No. 2018DM0034).
Computational ressources were provided by the J\"{u}lich Supercomputing Centre at the 
Forschungszentrum J\"{u}lich.

\appendix

\section{Density and current operators \label{app_operators}}

Here, we define the various nucleon density, current and derivative operators that we are using. 
Following Refs.~\cite{Borasoy:2006qn,Borasoy:2007vi,Alarcon:2017zcv}, we define the local 
density operators. The LO and NLO density operators include contact, contact isospin, contact 
spin as well as contact spin-isopsin operators given by:
\begin{align}
\rho(\vec{n}) &= \sum_{i,j = 0,1} a_{i, j}^\dagger(\vec n) a_{i, j}^{}(\vec n), \\
\rho_I^{}(\vec n) &= \sum_{i, j, j^\prime = 0,1} 
a_{i, j}^\dag(\vec{n}) (\tau_I^{})_{j, j^\prime}^{} a_{i, j^\prime}(\vec{n}), \\
\rho_S^{}(\vec n) &= \sum_{i, i^\prime, j = 0,1} 
a_{i, j^\prime}^\dag(\vec{n}) (\sigma_S^{})_{i, i^\prime}^{} a_{i^\prime, j} (\vec{n}), \\
\rho_{S,I}^{}(\vec n) &= \sum_{i, i^\prime, j, j^\prime = 0,1} 
a_{i, j}^\dag(\vec{n}) (\sigma_S^{})_{i, i^\prime}^{} (\tau_I^{})_{j, j^\prime}^{} a_{i^\prime, j^\prime}(\vec{n}),
\end{align}
while the current, isospin, spin and spin-isospin current density operator are given by 
\begin{align}
\begin{split}
\Pi_l^{}(\vec n) &= \sum_{i,j = 0,1} a_{i,j}^\dagger(\vec n) \nabla_l^{} a_{i,j}^{}(\vec n) \nn \\
&- \sum_{i,j = 0,1} \nabla_l^{} a_{i,j}^\dagger(\vec n) a_{i,j}^{}(\vec n),
\end{split}\\
\begin{split}
\Pi_{l, I}^{}(\vec n) &= \sum_{i,j,j^\prime = 0,1} 
a_{i,j}^\dagger(\vec n)(\tau_I^{})_{j,j^\prime}^{} \nabla_l^{} a_{i,j^\prime}^{}(\vec n) \nn \\ 
&- \sum_{i,j,j^\prime = 0,1} \nabla_l^{} a_{i,j}^\dagger(\vec n)(\tau_I^{})_{j,j^\prime}^{} a_{i,j^\prime}^{}(\vec n), 
\end{split}\\
\begin{split}
\Pi_{l,S}^{}(\vec n) &= \sum_{i,i^\prime,j = 0,1}
a_{i,j}^\dagger(\vec n)(\sigma_S^{})_{i,i^\prime}^{} \nabla_l^{} a_{i^\prime,j}^{}(\vec n) \nn \\
&- \sum_{i,i^\prime,j = 0,1} \nabla_l^{} a_{i,j}^\dagger(\vec n)(\sigma_S^{})_{i,i^\prime}^{} a_{i^\prime, j}^{}(\vec n),
\end{split}\\
\begin{split}
\Pi_{l,S,I}^{}(\vec n) &= \sum_{i,i^\prime,j,j^\prime = 0,1} 
a_{i,j}^\dagger(\vec n) (\sigma_S^{})_{i,i^\prime}^{} (\tau_I^{})_{j,j^\prime}^{} \nabla_l^{} a_{i^\prime, j^\prime}^{}(\vec n) \nn \\
&- \sum_{i,i^\prime,j,j^\prime = 0,1} \nabla_l^{} a_{i,j}^\dagger(\vec n) (\sigma_S^{})_{i,i^\prime}^{} (\tau_I^{})_{j,j^\prime}^{} a_{i^\prime,j^\prime}^{}(\vec n),
\end{split}
\end{align}
where the derivative operator reads $\nabla_l f(\vec{n})=(1/2)[f(\vec{n}+\hat{e}_l)-f(\vec{n}-\hat{e}_l]$. 
Due to the inclusion of electromagnetic corrections, we include isospin-up and -down operators
\begin{align}
\rho_p(\vec{n})&=a^\dagger(\vec{n})(\mathbb{1}+\tau_3)a(\vec{n}),\\
\rho_n(\vec{n})&=a^\dagger(\vec{n})(\mathbb{1}-\tau_3)a(\vec{n}),
\end{align}
as well.


\section{Uncertainty analysis \label{app_errors}}

From the definition of $\chi^2$ given in Eq.~(\ref{chisquare}), we note that 
$\chi^2$ is a function of the LO and NLO coupling constants
\begin{align}
\chi^2_{\rm LO} &= \chi^2\left(C_{^1S_0}, C_{^3S_1},\right),\\
\chi^2_{\rm N2LO} &= \chi^2\left( C_1^{}, \ldots,  C_{10}\right),\\
&\ldots \nonumber.
\end{align}
$\chi^2$ can be expanded around its minimum, giving 
\begin{equation}
\chi^2 = \chi^2_\mathrm{min} + \frac{1}{2} \sum_{i, j} h_{ij}^{} 
(C_i^{} - C_i^\mathrm{min})(C_j^{} - C_j^\mathrm{min}) + \ldots,
\end{equation}
where the Hessian matrix is denoted by
\begin{equation}
h_{ij}^{} \equiv \frac{\partial^2 \chi^2}{\partial C_i^{} \partial C_j^{}},
\end{equation}
and $C_i^\mathrm{min}$ denotes the result of the $\chi^2$~fit.
Then the error (or variance-covariance) matrix is defined as
\begin{equation}
\mathcal{E}_{ij}^{} \equiv \frac{1}{2} \big [ h_{}^{-1} \big ]_{ij},
\label{eq:covmatrix}
\end{equation}
while the standard deviations of the fit parameters read
\begin{equation}
\sigma_i^{} 
= \sqrt{\mathcal{E}_{ii}^{}}.
\label{uncertainty_coefficient}
\end{equation} 
Following Ref.~\cite{Alarcon:2017zcv} we again find a very large $\chi$ due to the underestimation 
of the PWA errors. Hence we have to rescale $\chi^2$ in the case of the phase shift calculation 
to get reasonable error estimates. In this case, we rescale $\chi^2$ by 
\begin{equation}
\chi^2 \rightarrow  N_\mathrm{dof}^{} \frac{\chi^2}{\chi^2_\mathrm{min}},
\end{equation}
such that $\chi^2/N_{\mathrm{dof}} \approx 1$ in the minimum \cite{birge,Perez:2014yla}.
\\
In the analysis, we also have to propagate the errors of the LECs to physical observables 
like phase shifts or binding energies. For a given observable $\mathcal{O}$, we assign an 
uncertainty according to
\begin{equation}
\Delta \mathcal{O} \equiv \sqrt{(J_\mathcal{O}^{T})_i^{} \mathcal{E}_{ij}^{} (J_\mathcal{O}^{})_j^{}}, 
\label{uncertainty_observable} 
\end{equation}
where 
\begin{equation}
(J_{\mathcal O})_i^{} \equiv \frac{\partial \mathcal{O}}{\partial C_i^{}},
\end{equation}
is the Jacobian vector of $\mathcal{O}$ with respect to the $C_i^{}$.


\end{document}